\documentclass{elsart}
\usepackage{graphics}
\usepackage{psfrag}
\usepackage{epsfig}
\usepackage{epsf}
\usepackage{float}
\usepackage{amssymb,stmaryrd,latexsym}
\newcommand{\be}{\begin{eqnarray}}
\newcommand{\ee}{\end{eqnarray}}
\usepackage{amsmath}
\usepackage{amssymb}
\usepackage{epsfig}

\begin{document}
\hfill{\small FZJ--IKP(TH)--2006--20, HISKP-TH-06-22}

\begin{frontmatter}
\title{Dispersive and absorptive corrections to the pion-deuteron scattering length}

\author{V. Lensky$^{1,2}$, V. Baru$^{2}$, J.~Haidenbauer$^1$, C.~Hanhart$^1$,}
\author{A. Kudryavtsev$^2$, and U.-G. Mei\ss ner$^{1,3}$}

{\small $^1$ Institut f\"{u}r Kernphysik, Forschungszentrum J\"{u}lich GmbH,}\\ 
{\small D--52425 J\"{u}lich, Germany} \\
%{\small and} \\
{\small $^2$ Institute of Theoretical and Experimental Physics,} \\
{\small 117259, B. Cheremushkinskaya 25, Moscow, Russia} \\
%{\small and} \\
{\small $^3$ Helmholtz-Institut f\"{u}r Strahlen- und Kernphysik (Theorie), 
} \\ 
{\small Universit\"at Bonn, Nu{\ss}allee 14-16, D--53115 Bonn, Germany 
}

\begin{abstract}
\noindent 
We present a parameter--free calculation of the dispersive and
 absorptive contributions to the pion--deuteron scattering length based
 on chiral perturbation theory. 
We show that once all diagrams
 contributing to leading order to this process are included, their net
 effect provides a   small correction to
 the real part of the pion--deuteron scattering length.
 At the same time the sizable imaginary part of the pion--deuteron
scattering length is reproduced accurately. 
\end{abstract}

\end{frontmatter}

{\bf 1.} The pion-nucleon ($\pi N$) scattering lengths are fundamental
quantities of low--energy hadron physics since they test the QCD
symmetries and the pattern of chiral symmetry breaking. As stressed by
Weinberg long time ago, chiral symmetry suppresses the isoscalar $\pi
N$ scattering length $a_+$ substantially compared to its isovector
counterpart $a_- \, $. Thus, a precise determination of $a_+$ demands
in general high accuracy experiments.

Here
pion-deuteron ($\pi d$) scattering near threshold plays an exceptional role for
$\mbox{Re}(a_{\pi d}) = 2a_+ + (\mbox{few--body corrections}) \ .$
 The first term $\sim a_+$
is simply generated from the impulse approximation (scattering off the proton
and off the neutron) and is independent of the deuteron structure.
Thus, if one is able to calculate the few--body corrections  in a controlled way, 
$\pi d$ scattering is a prime reaction to extract $a_+$
(most effectively in combination with pionic hydrogen measurements).
In addition, already at threshold the $\pi d$ scattering length is a
complex-valued quantity. It is therefore also important to gain a precise
understanding of its imaginary part - this is one of the issues addressed
in this letter.

Recently the $\pi d$ scattering length was measured to be \cite{PSI1}
\be
a_{\pi d}^{\mbox{exp}} =\left (-26.1\pm 0.5\mbox \, + \, i (6.3\pm 0.7)\right )\times
\,10^{-3} \ m_\pi^{-1} \ , 
\label{exp}
\ee 
where $m_\pi$ denotes the mass of the charged
pion. In the near future a new measurement with a projected total uncertainty of 0.5\% for
the real part and 4\% for the imaginary part of the scattering length will be
performed at PSI \cite{detlev}. Clearly, performing calculations up to this accuracy 
poses a challenge to theory that several groups recently took up
\cite{BBEMP,rus,doeringoset,arriola,mitandreas,danielneu}. In addition, an
interesting isospin violating effect in pionic deuterium was found, see \cite{MRR}.
 For a review on older
work we refer to Ref. \cite{al}.

The imaginary part for the $\pi d$ scattering length  can
be expressed  by unitarity in terms of the $\pi d$ total cross section through the
optical theorem. One gets
\begin{equation}
4\pi \mbox{Im}(a_{\pi d})=\lim_{q\to 0}q\left\{
\sigma(\pi d\to NN)+\sigma(\pi d\to \gamma NN)\right\} \ ,
\label{opttheo}
\end{equation}
where $q$ denotes the relative momentum of the initial $\pi d$ pair.
The ratio $R=\lim_{q\to 0}\left(\sigma(\pi d\to NN)/\sigma(\pi d\to
\gamma NN)\right)$ was measured to be $2.83\pm 0.04$
\cite{highland}. At low energies diagrams that lead to a sizable
imaginary part of some amplitude are expected to also contribute
significantly to its real part.
% due to
%the so--called principal value part of the corresponding
%integral.
 Those contributions are called dispersive corrections.  As a
first estimate Br\"uckner speculated that the real and imaginary part
of these contributions should be of the same order of magnitude
\cite{brueck}. This expectation was confirmed within Faddeev
calculations in Refs. \cite{at}.  Given the high accuracy of the
measurement and the size of the imaginary part of the scattering
length, another critical look at this result is called for as already
stressed in Refs. \cite{tle,bk}.  A consistent calculation is only
possible within a well defined effective field theory --- the first
calculation of this kind is provided here.

\begin{figure}[t!]
\begin{center}
\epsfig{file=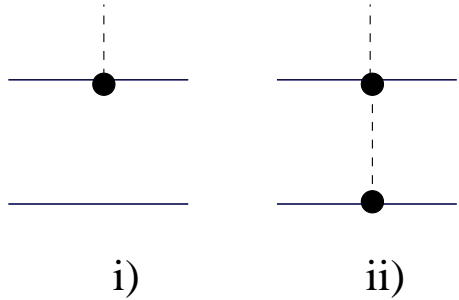, height=4cm, angle=0}
\caption{Diagrams contributing to the $\pi NN\to NN$ transition
up to the order considered:  {\it i)} direct contribution and  {\it ii)} rescattering.}
\label{kernel}
\end{center}
\end{figure}

What is needed a priori for such an endeavor is a controlled power
counting for $NN\to NN\pi$ using chiral perturbation theory (ChPT)
that is consistent with the one used for $\pi d$ scattering.  This was
developed in recent years \cite{cohen,hereweneedmany} --- for a review
we refer to Ref.~\cite{report}\footnote{For calculations that use
chiral perturbation theory in the standard formulation for the
reactions $NN\to NN\pi$ we refer to Refs. \cite{theothers}.}.  This
scheme led to the first %field theoretically consistent EFT
calculation for $pp\to d\pi^+$ \cite{ourpid}.  It was the central
finding of this work that all loops that contribute to $NN\to NN\pi$
at next--to--leading order cancel and the full transition amplitude up
to next--to--leading order solely contains the diagrams shown in
Fig. \ref{kernel}, however, with one modification compared to the
standard treatment as used already in Ref. \cite{koltunundreitan}: The
$\pi N\to \pi N$ vertex in diagram $ii)$ is to be used with its
on--shell value, $2m_\pi$, instead of the previously used value of
$3/2 \, m_\pi$. This increase was sufficient to bring the calculation
in agreement with the data.

In addition to the hadronic part of the dispersive and absorptive corrections to
 the $\pi d$ scattering length,
 we estimate the corresponding contribution from the transition $\pi d\to \gamma NN\to \pi d$
using the full structure of the one--photon exchange.  Note that 
the inelastic channel $\pi d\to \gamma NN$ accounts for $1/4$ of the imaginary
part of $a_{\pi d}$ and therefore one can expect a sizable
contribution also to its real part. 

%In this paper we present a first chiral perturbation theory calculation for
%the leading dispersive corrections.
% Our work provides not only a reliable
%estimate for this contribution and its uncertainty, but at the same time
%explains the differences to previous works.

The paper is organized as follows: in the next section we will present
the power counting for the $\pi d$ system including the dispersive part.
In the third section we give our results, while a comparison to previous works 
is done in Sec.~4. We close with a brief summary.

\begin{figure}[t!]
\begin{center}
\epsfig{file=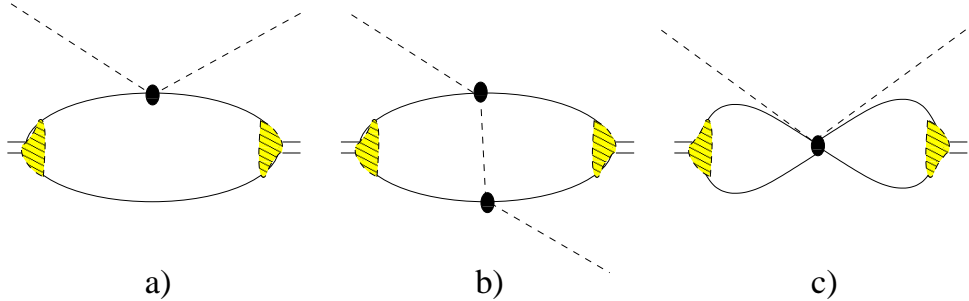, height=4cm, angle=0}
\caption{Typical Feynman diagrams for $\pi d$ scattering; shown are the one--body
 term (a), the double--scattering  contribution (b)
and a four--nucleon--contact term (c). Solid
  black dots stand for the $\pi N$ interaction, whereas the hatched area shows
  the deuteron wave function.}
\label{dia}
\end{center}
\end{figure}

{\bf 2.} The basis of any effective field theory calculation is a
proper power counting that allows one to organize the diagrams
according to some a priori known hierarchy.  It was argued by Weinberg
\cite{wein} that in case of pion reactions on few--nucleon systems and
especially on nuclei the Goldstone theorem ensures that one can expand
the transition operators perturbatively. Then those have to be
convoluted with proper wave functions. The important thing is to
identify the relevant expansion parameter in the transition
operators. For $\pi d$ scattering in \cite{wein} the series is
organized in powers of momenta and pion masses in units of the chiral
symmetry breaking scale $\Lambda_\chi\sim 1$~GeV. The typical one--
and two--body diagrams are shown in Fig \ref{dia} (a) and (b),
respectively.  The small binding energy of the deuteron introduces a
new small scale that can be accounted for systematically \cite{BBEMP}.
In Refs. \cite{recoils1,recoils2} it was demonstrated how the scheme
is to be modified in the presence of three--body ($\pi NN$) cuts.
Based on calculations with deuteron wave functions obtained solely
from contact $NN$ interactions, in Refs. \cite{gries,silasmartin} it
was argued that field theoretical consistency calls for a counter term
at leading order. However, in
Refs.~\cite{arriola,mitandreas,danielneu} it was shown that this is no
longer necessary as soon as the finite range of the one--pion exchange
is included in the $NN$ potential.

\begin{figure}[t!]
\begin{center}
\epsfig{file=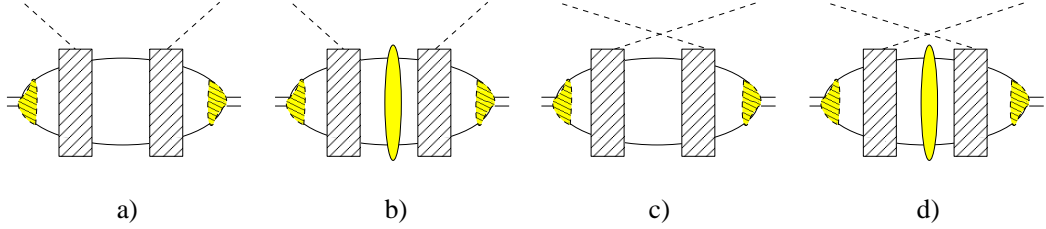, height=3cm, angle=0}
\caption{The classes of the hadronic contributions to the dispersive
corrections. Diagram (a) and (b) denote the direct
and (c) and (d) the crossed terms.  The filled ellipse denotes
the $NN$ interaction in the intermediate state.  The diagrams
contributing to the $\pi NN\to NN$ transition to the given order are
shown in Fig. \ref{kernel}. The diagrams with emission of pion on the
second nucleon are not shown explicitly but taken into account in the
calculation. }
\label{disp}
\end{center}
\end{figure}

\begin{figure}[t!]
\begin{center}
\epsfig{file=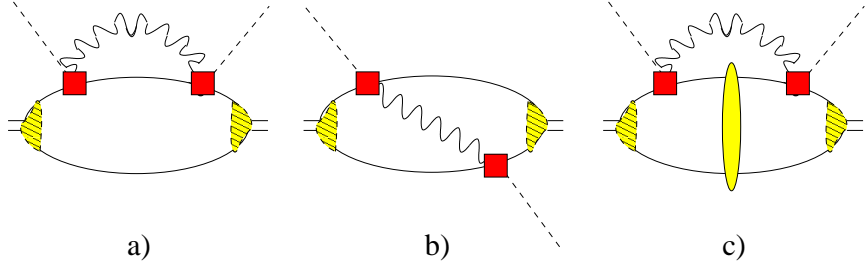, height=3.5cm, angle=0}
\caption{Diagrams contributing to the
dispersive corrections from photon--exchange interactions:
 one--body term (a) and double scattering (b). The filled
ellipse denotes the $NN$ interaction in the intermediate state.
The solid squares denote the full $\pi N\to \gamma N$ transition 
amplitude as depicted in Fig. \ref{redbox}.}
\label{photon}
\end{center}
\end{figure}

So far no attempt was made to also include consistently --- within
ChPT ---  the so called dispersive corrections that emerge from the
hadronic $\pi d\to NN\to \pi d$  and photonic $\pi d\to \gamma NN\to \pi d$ 
reaction chains. We define dispersive corrections as contributions from diagrams 
with an intermediate state that contains only nucleons, photons and at most real pions.
%According to our definition the diagrams that contribute to the dispersive corrections 
%should have intermediate states that contain only nucleons, photons and at most real pions. 
Thus, all other potentially important 
corrections to the $\pi d$ scattering length that come from, e.g., 
the $\Delta$ resonance   will not be discussed in this letter. 
The diagrams contributing to the hadronic and photonic parts of the 
dispersive corrections in accordance to our definition are shown schematically 
in Figs. \ref{disp} and \ref{photon}, respectively.

 Before we present the results
of the calculation we first need to establish the power counting. The
fact that the hadronic reaction chain  $\pi d\to NN\to \pi d$ is a process with large momentum
transfer introduces a new scale into the problem that needs to be
accounted for by a modified power counting.

%So far no attempt was made to also include consistently --- within
%ChPT --- the so called dispersive corrections that emerge from the
%reaction chain $\pi d\to NN\to \pi d$. Before we present the results
%of the calculation we first need to establish the power counting. The
%fact that this reaction chain is a process with large momentum
%transfer introduces a new scale into the problem that needs to be
%accounted for by a modified power counting.

%
%\begin{figure}[t!]
%\begin{center}
%\epsfig{file=disp.eps, height=8cm, angle=0}
%\caption{The two classes of the hadronic
%contribution to the dispersive corrections
%without (with) inclusion of the $NN$ interaction in the 
%intermediate state: direct (a) ((c)) and 
%crossed (b) ((d)).
%The filled ellipse denotes the $NN$ interaction in the intermediate
%state.
% The diagrams contributing to the $\pi NN\to NN$ transition
%to the given order are shown in Fig. \ref{kernel}. The diagrams with emission of pion on the
%second nucleon are not shown explicitly but taken into account in the calculation. } 
%\label{disp}
%\end{center}
%\end{figure}
%

To establish the counting scheme we focus on two--body currents only
--- how to include one--body currents into the standard scheme is
described in Ref. \cite{wein}\footnote{\label{onebodyprod} In Ref.
\cite{cohen} the corresponding recipe is given that needs to be
applied to $NN\to NN\pi$ and therefore also to the dispersive
corrections. It implies that the diagrams shown in
Fig. \ref{kernel}(a) and (b) contribute at the same order for
$s$--wave pion production.}.  Thus, in what follows we will compare our
two body $\pi NN\to NN\to \pi NN$ operators with the leading two body
operator shown in Fig. \ref{dia}(b).  Then it is sufficient to read
off the vertex factors for the $\pi NN\to \pi NN$ transitions to
identify the order of any given diagram.  We therefore estimate
$m_{\pi}^2/(f_{\pi}^4\: q^2 )$ for the diagram (b) of
Fig. \ref{dia}  where $q$ here defines the momentum of intermediate pion. 
Utilizing Weinbergs counting scheme where all internal
momenta are assumed to be of order $m_\pi$ we find diagram (b)
% Using this method we find diagram (b) 
to be
$\mathcal{O}(1)$ --- here and in what follows we drop a factor $1/f_\pi^4$ common to all
diagrams to get the order estimate.  Power counting gives that the $4N2\pi$ 
contact term shown in Fig.\ref{dia} (c)
contributes at $\mathcal{O}(\chi^2)$, where $\chi=m_\pi/M_N$ is the
standard expansion parameter of ChPT with $M_N$ for the
the nucleon mass. This last contribution comes with a yet
unknown coefficient. As such, an estimate for its size provides the
theoretical accuracy that a calculation for the $\pi d$ scattering length 
can have at most.
Therefore, assuming naturalness for the strength of
the contact term, the theoretical limit of accuracy is of order $\chi^2$ which
translates into a few
percent.  Reverting this statement, in order to reach a theoretical accuracy
that is comparable to that expected for the experimental value of the $\pi d$
scattering length, all contributions of lower order than  $\chi^2$ should be evaluated.
We will now show that the dispersive corrections contribute to
$\mathcal{O}(\chi^{3/2})$.

Transitions of the type $\pi NN\to NN\to \pi NN$ --- sketched 
in Fig. \ref{disp} (a) --- get
contributions from small values of the $NN$ intermediate momentum $q$
($q\sim m_\pi$ or smaller) as well as from large values of $q$ ($q\sim
p_{\mbox{thr}}= \sqrt{m_\pi M_N}$). The latter value refers to the
on--shell momentum of the intermediate $NN$ state (or, equivalently,
to the threshold initial momentum of the reaction $NN\to NN\pi$). The
power counting as given for $NN\to NN\pi$ relates to the latter part
of the contribution. It is based on the assignment
$p_{\mbox{thr}}/M_N\sim \mathcal{O}(\chi^{1/2})$ \cite{cohen}.  To
derive the order of the dispersive corrections let us start with
diagram $d1$ of Table \ref{results}.  (See footnote \ref{onebodyprod} for how to 
include diagrams of the type $d2$). 
For this one we find in units of 
the amplitude for diagram (b) of Fig. \ref{dia} (estimated to be $\mathcal{O}(1)$)
\footnote{For a brief description on how to identify the order of
a particular diagram we refer to Appendix E of Ref. \cite{report}.}

\vspace{-0.3cm}

\be \hspace*{-0.7cm}\left[\left.\left(\frac{m_\pi}{f_\pi^3 q}\right)^2 \frac1{m_\pi-q^2/M_N+i\epsilon}
\left(\frac{q^3}{(4\pi)^2}\right)\right]\right/\left(\frac1{f_\pi^4}\right) \sim \left\{\begin{array}{ll}
\mathcal{O}\left(\chi^2\right) & \mbox{for} \, q\sim m_\pi \\
\mathcal{O}\left(\chi^\frac32\right) & \mbox{for} \, q\sim
p_{\mbox{thr}} \\
\end{array}
\right.
\label{count}
\ee where the first term in the square brackets comes from the $\pi NN\to NN$
transition operator, the second one from the $NN$ propagator in the
intermediate state and the last one from the integral measure. 
%To arrive at this order of magnitude estimate
%when the two nucleon intermediate state is almost on shell, i.e. for $q\sim p_{\mbox{thr}}$,
%we substituted the two nucleon propagator by its $\delta$-function term for 
%the size estimate should apply to both the real
%part as well as the imaginary part that emerges from $\pi NN\to NN\to \pi NN$. 
To arrive at the estimate for $q\sim p_{\mbox{thr}}$, where the l.h.s of 
Eq.(\ref{count})  involves a singularity, we replaced the two--nucleon propagator
 by the corresponding  $\delta$-function term for this estimate should apply 
to both the real part as well as the imaginary part that emerges from $\pi NN\to NN\to \pi NN$. 
Furthermore we used $4\pi f_\pi\simeq M_N$.
The small momentum part of the integral is thus of order
$\chi^2$ and not relevant for this study.  That is why dispersive
corrections were not considered in the studies of Refs.
\cite{BBEMP,rus}.  However, the part of the integral where $q$ is of the order of 
$p_{\mbox{thr}}$ is indeed of lower order than $\chi^2$ and thus should
be considered.  It is important to stress that
for a consistent understanding of $NN\to NN\pi$ within ChPT
it was also necessary to include
the large scale $p_{\mbox{thr}}$ explicitly in the power counting
\cite{cohen,report}.  
%The size estimate should apply to both the real 
%part as well as the imaginary part that emerges from $\pi NN\to NN\to \pi NN$. 
For the imaginary part of the amplitude $\pi NN\to NN\to \pi NN$  we have an experimental value ---
(3/4)Im$(a_{\pi d})$/Re$(a_{\pi d}) \simeq 1/6$, where the factor of 3/4
was introduced since this fraction  of the width comes from $\pi d\to NN$.
To check  the power counting we need some
estimate for the real part of the scattering length, which is known to
be dominated by the double rescattering term (Fig. \ref{dia}(b)) and
was shown above to be $\mathcal{O}(1)$. Therefore we expect from the
above considerations a relative suppression of the imaginary part to
the real part of the order of $(m_\pi/M_N)^{3/2}=1/17$.  Thus, the hadronic
contribution to the imaginary part of the $\pi d$ scattering length is
about a factor of 3 larger than predicted by the power counting
 --- a deviation that is
tolerable.

Since the $NN$ interaction is non--perturbative in
diagrams, where a two--nucleon state contributes near on--shell, the
full $NN$ interaction should be taken into account --- the
corresponding diagram is depicted in Fig. \ref{disp}(b).  These are
also included in our calculation using the techniques outlined in
Ref. \cite{recoils2}.  At the same chiral order there is an additional
class of contributions --- these are the crossed terms shown as
diagrams (c) and (d) in Fig. \ref{disp}.  Already in
Ref. \cite{thouless} an evaluation of those diagrams was called for,
however, a consistent calculation of the terms that emerge from diagram
$ii)$ of Fig. \ref{kernel} has not been possible so far \cite{al}.  Since
 we work within a consistent field theory no such problem exists.
The numerical importance of some crossed diagrams for $\pi d$
scattering was already observed before and referred to as the
so-called Jennings mechanism \cite{jen}: to understand data measured
with tensor polarized deuterons for elastic $\pi d$ scattering at
backward angles, a particular crossed pion contribution needs to be
included --- see also the discussion in Ref. \cite{garmiz}.  The order
assignment for the crossed diagrams is obvious once one applies the
same procedure that leads to the estimate given in (\ref{count}) ---
the only necessary change is to switch the sign of $m_\pi$ in the $NN$
propagator.  Note, in these diagrams the two--nucleon intermediate
state is always off--shell in contrast to the $NN$ state in the direct
contributions that are expected to receive the dominant contributions
from (near) on--shell nucleons. It is therefore surprising that direct
and crossed terms appear at the same order.  However, one should
recall that the chiral expansion is an expansion around the chiral
limit. When approaching the chiral limit the intermediate two--nucleon
states of both direct and crossed diagrams approach the same
kinematical point. Therefore it is natural that they also contribute
to the same chiral order for physical pion masses.

A priori there is no rule how to include the electromagnetic contribution to
both the real and the imaginary part of the $\pi d$ scattering length (see
Fig. \ref{photon}) into the
power counting --- the fine structure constant $\alpha$ is clearly an
independent parameter. Based on the observation that the electromagnetic
and the hadronic contribution to the imaginary part are of the same order of
magnitude, we assign the same chiral order to both --- as it is often done
in chiral perturbation theory studies.

\begin{figure}[t!]
\begin{center}
\epsfig{file=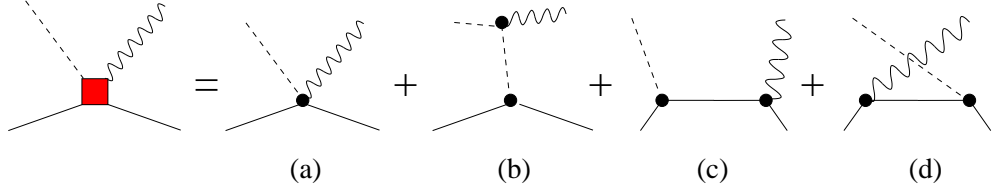, height=2.5cm, angle=0}
\caption{Diagrams contributing to the
$\pi N\to\gamma N$ transition operator:
Kroll--Rudermann term (a), pion in flight (b) and nucleon s-- (c) and
u-- (d) channel.}
\label{redbox}
\end{center}
\end{figure}

{\bf 3.} In this section we present the results of our investigations. 
We first focus on the hadronic contribution to the dispersive corrections.
The results are given in Table \ref{results}.
Note, for the contributions that involve the $NN$ interaction
in the intermediate state we do not give the individual
results explicitly.  All calculations are done with  the CD-Bonn  
NN potential \cite{Machleidt}.

First of all we observe that with a value of $4.25 \times 10^{-3} \
m_\pi^{-1}$ the imaginary part of the scattering length turns out to
be very close to the experimental number of $(3/4)$Im$\left(a_{\pi
d}^{\mbox{exp}}\right)=(4.7\pm 0.5)\times 10^{-3} \ m_\pi^{-1}$
(c.f. Eqs.  (\ref{exp}) and (\ref{opttheo}) and discussion below the
latter). This is not too surprising given the good description for the
near threshold cross section of $NN\to d\pi$ as reported in Ref.
\cite{ourpid}. For the real part on the other hand we observe a
striking pattern: although the individual contributions can be quite
large, the total sum turns out to be  rather small.  About 1/3 of the
contribution from the direct diagrams of Table \ref{results} gets
canceled by the corresponding ones with $NN$ interactions in the
intermediate state. This is the same pattern as for the imaginary part
--- the fact that the latter reduction in the magnitude is natural for
processes of the type $NN\to NNx$ was discussed in
Ref. \cite{withkanzo}.  However,  about 60 \% of the contribution of the
direct diagrams of Table \ref{results} is canceled by the
corresponding crossed diagrams.  As we will see, part of this
cancellation is quite natural. When comparing the direct and the
crossed diagrams we observe that the two--nucleon propagator in the
intermediate state of the former reads $1/(m_\pi-q^2/M_N)$ (c.f. Eq.
(\ref{count})), where $q$ denotes the relevant loop momentum.  The
corresponding expression for the latter reads
$1/(-m_\pi-q^2/M_N)$. Thus, for small values of $q$, where the
two--nucleon propagator becomes static, one obtains $1/m_\pi$ and $-1/m_\pi$
respectively, and some contributions of direct and crossed diagrams, specifically d2 and  c2,
will largely cancel. 
%{\bf In principle, the same arguments also apply to  other diagrams. However
%since we include the full TOPT structure of the pion propagator only part of direct 
%diagrams with pion exchange will cancel at small momenta the corresponding contributions 
%of crossed diagrams.}
Note, this  cancellation does not mean that the full contribution of each
pair of diagrams cancels.  The numbers given in the table also contain
the contributions from large values of $q$, where such a cancellation
does not necessarily occur.  Note also that it is the  structure of the two nucleon propagator 
that is responsible for the smallness of diagram d1 as compared to its crossed partner c1. 
In contrast to diagram   c1, the two--nucleon propagator in d1 changes its sign when passing through the point 
$q^2=m_{\pi}M_N$ whereas all  other terms such as vertex functions, pion propagators, etc.,
have the same sign throughout the region of integration. Thus, a strong cancellation  takes 
place for the latter and as a result the real part of diagram d1 is much smaller than that of c1.
The situation for diagram d3 is different to d1, since also the S-wave 
deuteron wave function changes sign at $q\sim \sqrt{m_\pi M_N}$. This leads to a constructive 
interference of contributions from small and large momenta.

In Fig. \ref{photon} we show the diagrams that contribute to the
electromagnetic piece of the dispersive contributions.  To evaluate
the real part of the one--body contribution (diagram (a)) we use the
same prescription as used in Ref. \cite{recoils1}, namely we subtract
the term corresponding to the one--body operator at zero
momentum. This removes the leading divergence that in a full
calculation needs to be absorbed into the electromagnetic corrections
of the $\pi N$ scattering lengths (note, this quantity was recently
estimated in Ref. \cite{ericson})\footnote{ A more precise calculation
within QCD+QED requires a much more sophisticated framework, see
e.g. \cite{GRS} --- this is beyond the scope of this letter.}.  To
ensure gauge invariance the $\pi N\to \gamma N$ amplitude needs to
contain all diagrams shown in Fig. \ref{redbox}.   However, since for the calculation
of the photonic part of the dispersive correction the $\pi N\to \gamma N$ amplitude 
contributes at threshold, diagrams (c) and (d) are suppressed by one power of 
$\chi$ as compared to  diagrams (a) and (b). 
%since diagram (c) and (d) are suppressed by one power of $\chi$ for $s$--wave pions, 
Thus, we did not take them into account in this leading
order calculation. Because of the same reasoning we neglect also the contributions to the 
$\pi N\to \gamma N$ amplitude from the $\Delta$ resonance \cite{Fearing}.
We calculated the full $\pi d\to \gamma NN\to \pi
d$ amplitude in Coulomb gauge. The corresponding propagator is given,
e.g., in Ref. \cite{landaulifshitz}.  The final result is $(-0.1+i\,
1.4)\times 10^{-3} \ m_\pi^{-1} $. 
 Again, the imaginary part\footnote{Note that the photonic part of the absorption correction is dominated 
by the imaginary part of the $\pi^- p$ scattering length due to the process $\pi^- p \to \gamma n\to \pi^- p$.}
 is sufficiently close to the corresponding experimental  value of 
\begin{table}[H]
\begin{center}
\caption{Hadronic contribution to the real and imaginary part 
of $a_{\pi d}$ in units of $m_\pi^{-1} \times 10^{-3}$.
We only show the typical topologies --- all permutations
are included as indicated by the ellipses. Line code as in Fig. \ref{dia}.}
\vskip 0.33cm
\begin{tabular}{|c| l c r c l|}
\multicolumn{6}{l}{\it direct contributions} \\[0.2em]
\hline
$d1$ & \parbox[c]{7.5cm}{\epsfig{file=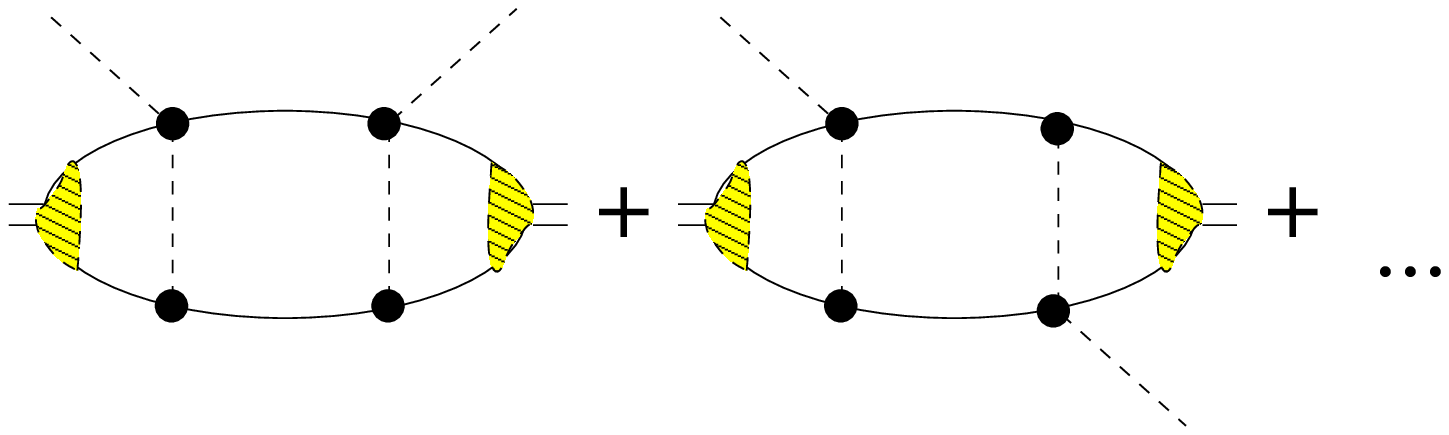, height=2.0cm}} & $=$ & $0.49$ & $+$ & $i \, 6.68$ \\ 
$d2$ & \parbox[c]{5cm}{\epsfig{file=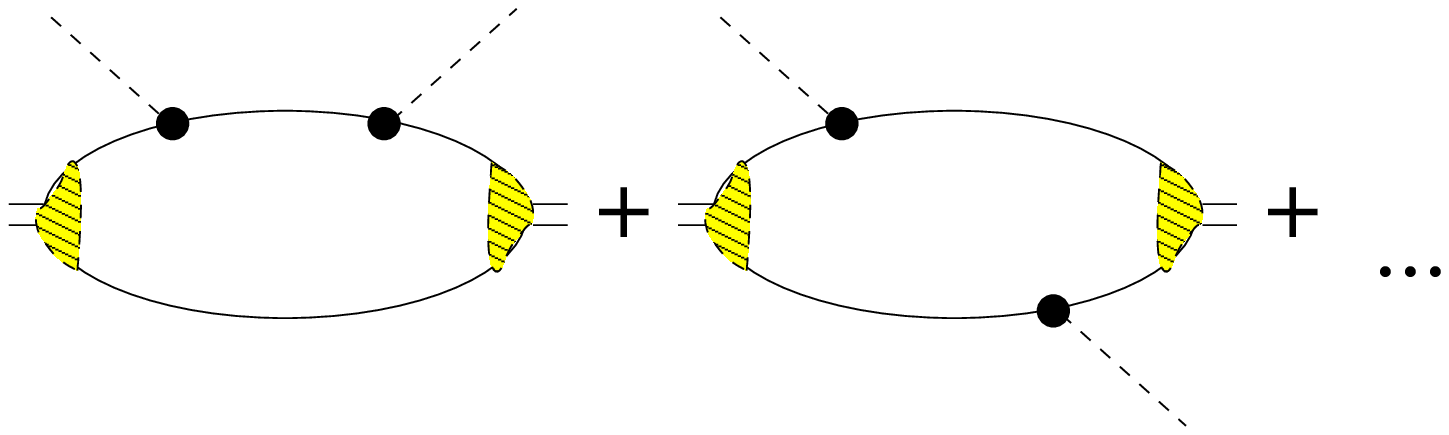, height=2.0cm}} & $=$ & $-0.8$ & $+$ & $i \, 0.1$ \\ 
$d3$ & \parbox[c]{5cm}{\epsfig{file=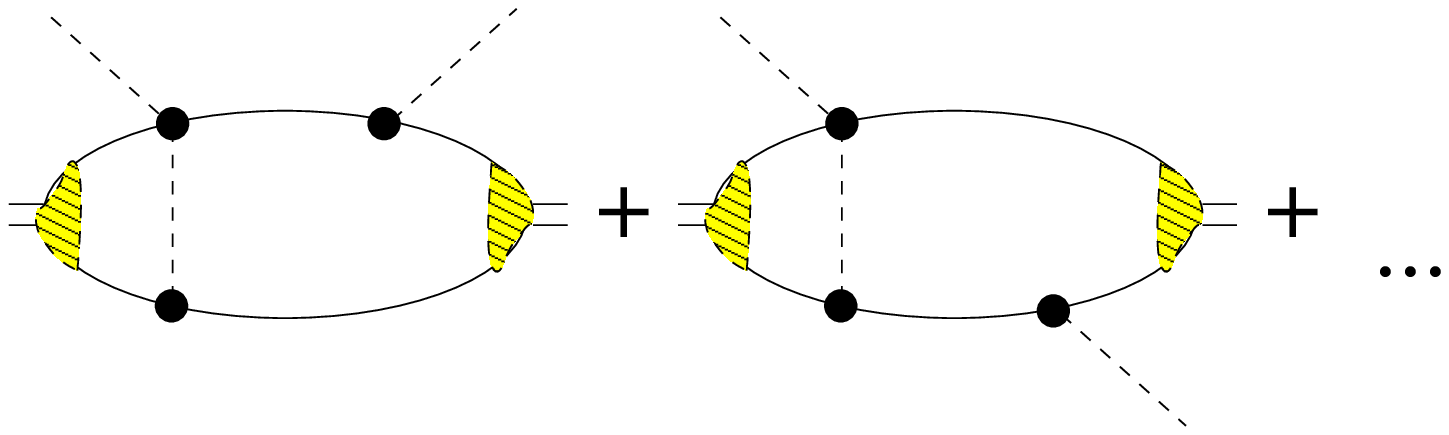, height=2.0cm}} & $=$ & $-6.02$ & $-$ & $i \, 1.65$ \\
\hline 
& sum of this group & $=$ & $-6.33$ & $+$ & $i \, 5.13$ \\ 
$d4{-}d6$ &  corresponding terms with intermediate $NN$ interaction & $=$ & $1.96$ & $-$ & $i\, 0.88$ \\ 
\hline 
& sum of all direct terms & $=$ & $-4.37$ & $+$ & $i \, 4.25$ \\
\hline
\multicolumn{6}{l}{\it crossed contributions} \\[0.2em]
\hline
$c1$ & \parbox[c]{5cm}{\epsfig{file=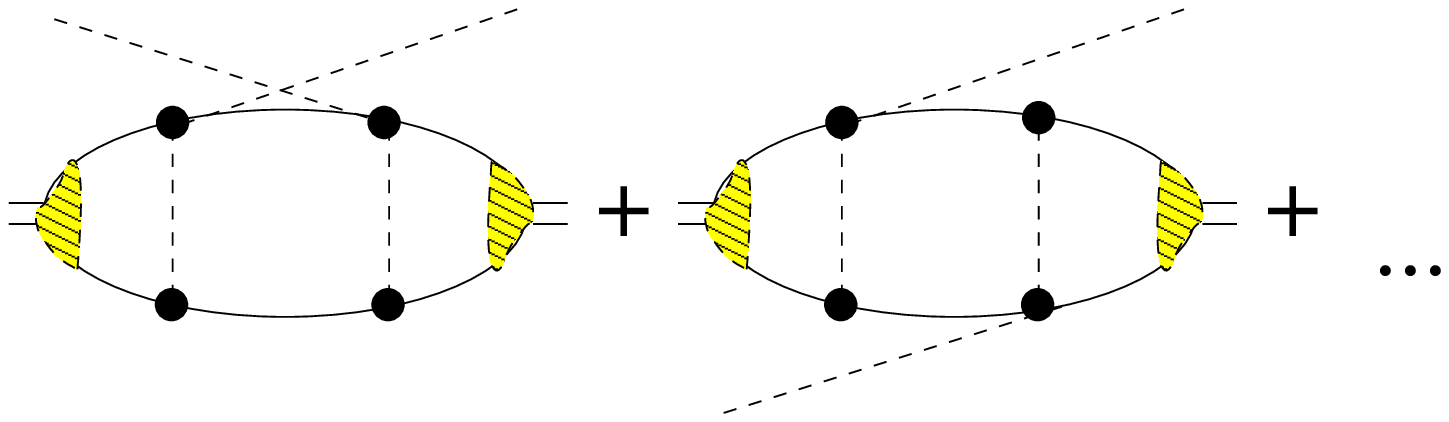, height=2.0cm}} & $=$ & $1.83$ & & \\ 
$c2$ & \parbox[c]{5cm}{\epsfig{file=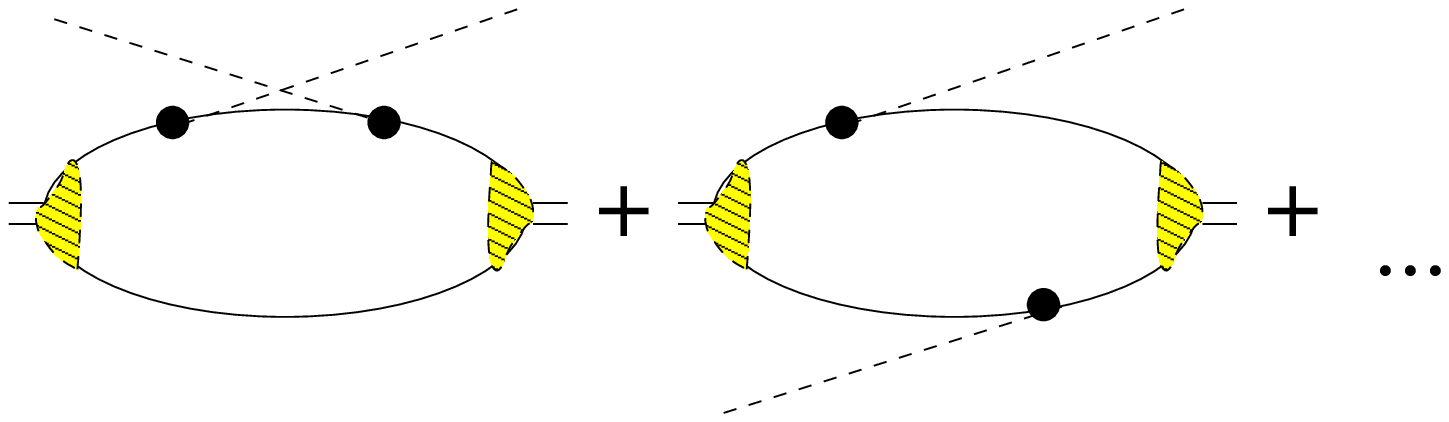, height=2.0cm}} & $=$ & $1.37$ & & \\ 
$c3$ & \parbox[c]{5cm}{\epsfig{file=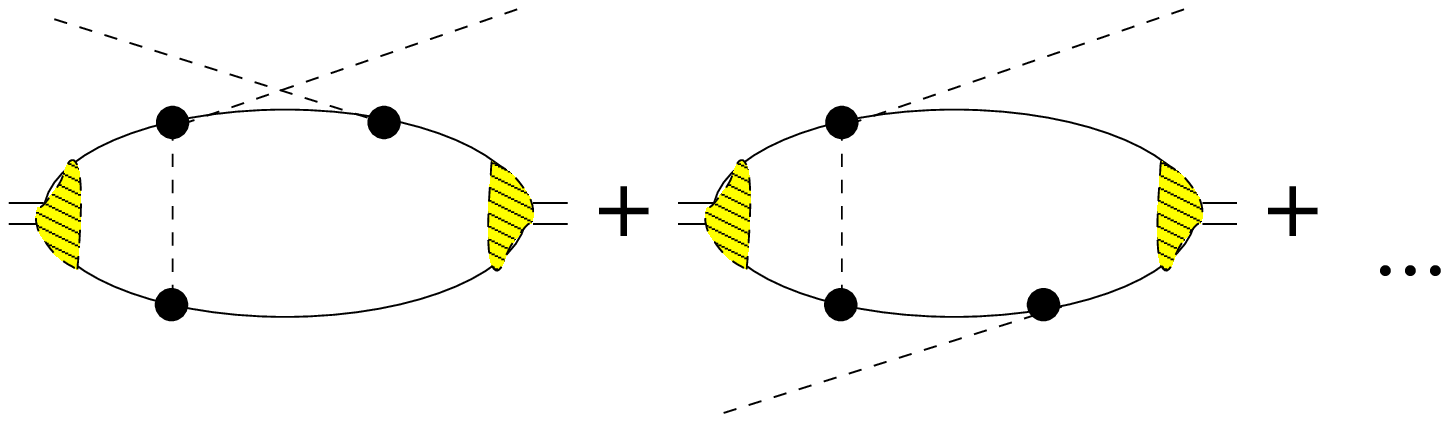, height=2.0cm}} & $=$ & $-0.09$ & &  \\
\hline 
& sum of this group & $=$ & $3.11$ & & \\ 
$c4{-}c6$ &   corresponding terms with intermediate $NN$ interaction  & $=$ & $-0.37$ & & \\ 
\hline 
& sum of all crossed terms & $=$ & $2.74$ & & \\
\hline 
\multicolumn{6}{l}{\it direct + crossed} \\[0.2em]
\hline 
& total sum  & $=$ & $-1.63$ & $+$ & $i\, 4.25$ \\ 
\hline
\end{tabular}
\label{results}
\vspace{0.3cm}
\end{center}
\end{table}
$(1/4)$Im$\left(a_{\pi d}^{\mbox{exp}}\right)=(1.6\pm 0.2)\times
10^{-3} \ m_\pi^{-1}$, whereas the real part is negligible. This
conclusion is consistent with that of Ref. \cite{tle}, where the real
part was found to be $-0.2\times 10^{-3} \ m_\pi^{-1}$ within a less
complete calculation\footnote{In Ref. \cite{rockmore} the
corresponding integral was evaluated to be one order of magnitude
larger; however, the author assumes the photon energy $q_0$ in the
exchange to vanish instead of using the correct value of $q_0\simeq
m_\pi$.}.

{\bf 4.} We now compare our results for the hadronic
contributions to other works. 
The dispersive corrections to $\pi d$ scattering were investigated
using Faddeev calculations in Refs. \cite{at}. Since these works
considered only those intermediate states that contain at most one
pion at a time, all direct diagrams were included.
The authors found $-5.6(1.4)\times  10^{-3} \, m_\pi^{-1}$
as contribution to the real part --- this number was also
used in the reanalysis of $\pi d$ scattering in Ref. \cite{tle}. This is to be compared with
$-4.37\times 10^{-3} \, m_\pi^{-1}$ from our work, which agrees to the above
result within the uncertainty. Note, the Faddeev equations
produce amplitudes that are non--perturbative in the $\pi N$ and the $NN$
interaction simultaneously. Thus, in addition to the direct terms
as shown in Table \ref{results} there are many more diagrams
included like contributions where the 
$NN$--pair interacts while there is a pion in flight. However, all
those are of higher order in the chiral expansion. The
closeness of our result for the direct terms and the corresponding
result of Ref. \cite{at} is thus an indication for the convergence of
the chiral expansion.  Recall that our final result to the real part of 
$a_{\pi d}$ is  smaller because of a cancellation of the mentioned contribution
with the crossed diagrams that were not included in Refs. \cite{at}.

In Ref. \cite{doeringoset} the diagrams $d1$ of Table \ref{results} were
evaluated explicitly, besides many others that are difficult to
match to our amplitudes. In this work they contribute with
$(0.24+i6.0)\times 10^{-3} \, m_\pi^{-1}$ to the $\pi d$ scattering
length.  This number is to be compared to our $(0.49+i6.68)\times
10^{-3} \, m_\pi^{-1}$.  In Ref. \cite{doeringoset} a value of $3/2
m_\pi$ was used at the $\pi N\to \pi N$ vertex in contrast to the
proper $2m_\pi$ as derived recently \cite{ourpid}. On the other hand,
Ref. \cite{doeringoset} 
includes a (small) isoscalar piece into this
vertex 
 --- chiral power counting assigns a subleading order to this
piece. However, the pattern of the result is the same: a
small real part is accompanied by a sizable imaginary part.

Our results for diagrams $d2$ and $c2$ agree to those of Ref. \cite{tbk},
once the same $\pi NN$ coupling constant is used.

Some of the diagrams in Table 1 where included in the phenomenological
studies of Refs.  \cite{doeringoset,bk}. In particular, the second
diagrams of d3 and c3 contribute to the so--called SP-interference
term (a double scattering diagram, where the first $\pi N\to \pi N$
transition is in an $s$--wave, whereas the second is $p$--wave). In
those studies the $p$--wave amplitude was taken from fits to $\pi N$
data and parameterized as a strength parameter times the square of the
$\pi N$ relative momentum.  Indeed, what appears at the lower nucleon can
be regarded as the $\pi N$ scattering in $p$--wave but in the boosted
frame.  However, this treatment misses an important momentum
dependence, since in the boosted frame the nucleon propagator in the
intermediate state contains a term $q^2/M_N$ in the denominator (see
Appendix). As a consequence the $p$--wave subamplitude in the
phenomenological studies grows quadratically with momentum, even for
momenta $q^2\sim m_{\pi}M_N$ and the full matrix element scales with
the wave function at the origin \cite{doeringoset}, which is
theoretically not controlled.  On the other hand, in our case this
subamplitude goes to a constant leading to a controlled behaviour of
the matrix element \cite{mitandreas}. This is why those parts of
Refs. \cite{doeringoset,bk} can not be matched to our
results.  Based on the arguments given, we call for microscopic
calculations, now possible within ChPT, instead of applying the phenomenological techniques.

{\bf 5.} To summarize, in this work we calculated for the first time
the absorptive and dispersive corrections to the $\pi d$ scattering
length using ChPT.  Especially we found for the absorptive part 
\be
\mbox{Im}(a_{\pi d}) =((4.25\pm 1.2)+(1.4\pm 0.4 ))\times \,10^{-3} \
m_\pi^{-1} \ 
\label{ours}
  \ee
 to be compared with the experimental value of
\be
\mbox{Im}\left(a_{\pi d}^{\mbox{exp}}\right) = 
((4.7\pm 0.5)+(1.6\pm 0.2))\times \,10^{-3} \ m_\pi^{-1} \ .
\label{expres}
\ee In both Eq. (\ref{ours}) as well as Eq. (\ref{expres}) we give the
hadronic and electromagnetic contribution separately. 
We thus find good agreement between theory and experiment for
each of the contributions.
 The theoretical
uncertainty is estimated to be of order $2\, m_\pi/M_N$ in both
cases\footnote{The factor of 2 appears, because the $\pi NN\to NN$ and
$\pi NN\to \gamma NN$ transition operators --- both evaluated with an
uncertainty of order $m_\pi/M_N$ --- appear twice in each amplitude.}.

For the corresponding dispersive part we get
\begin{equation}
a_{\pi d}^{disp}=-1.7\times 10^{-3} \, m_\pi^{-1} %\qquad \mbox{which gives} \qquad 
\Longrightarrow
a_{\pi d}^{disp}/\mbox{Re}\left(a_{\pi d}^{\mbox{exp}}\right) \sim 6.5\% \ .
\end{equation}
The number given for $a_{\pi d}^{disp}$ now contains both the hadronic
as well as the electromagnetic contribution and for Re($a_{\pi d}^{\mbox{exp}}$) we
used Eq. (\ref{exp}). This result is quite small given that the
imaginary part of the $\pi d$ scattering length is about 1/4 of the
real part. 
It is difficult to
provide a proper estimate for the uncertainty of $a_{\pi d}^{disp}$,
since it emerged from a cancellation of individually sizable
terms. The most naive method would be to use the uncertainty of order $2\, m_\pi/M_N$
 one has for, e.g., the sum of all direct diagrams
to derive a $\Delta a_{\pi d}^{disp}$ of around $1.4\times
10^{-3} \, m_\pi^{-1}$, which corresponds to about 6\% of
$\mbox{Re}\left(a_{\pi d}^{\mbox{exp}}\right)$. However, given that
the operators that contribute to both direct and crossed diagrams are
almost the same (see Appendix) and that part of the mentioned
cancellations is a direct consequence of kinematics, this number for
$\Delta a_{\pi d}^{disp}$ is probably too large. A reliable
error estimate for $a_{\pi d}^{disp}$ requires an explicit evaluation of
the next order contributions.

  We showed that the
smallness of the dispersive contribution to
the real part of the $\pi d$ scattering length is  a consequence of
efficient cancellations amongst various, individually sizable terms.
To gain a better understanding of the real part of the pion-deuteron
scattering length, a complete calculation of all isospin-breaking corrections
at N(N)LO in the EFT with virtual photons is called for (as also stressed
in \cite{MRR}).

\noindent 
{\bf Acknowledgments}

\noindent 
Useful discussions with J.~Gasser, D.~Gotta, E.~Oset, D.R.~Phillips, A.~Rusetsky, A.W.~Thomas,
and W.~Weise 
are gratefully acknowledged.
A.~K. and V.~B. also thank T.E.O.~Ericson for useful discussions.
This research is part of the EU Integrated Infrastructure Initiative
Hadron Physics Project under contract number RII3-CT-2004-506078, and
was supported also by the DFG-RFBR grant no. 05-02-04012 (436 RUS
113/820/0-1(R)) and the DFG SFB/TR 16 "Subnuclear Structure of Matter".  A.~K. and
V.~B. acknowledge the support of the Federal Program of the Russian
Ministry of Industry, Science, and Technology No 02.434.11.7091.

\renewcommand{\theequation}{A-\arabic{equation}}
\section*{Appendix}  % use *-form to suppress numbering
\setcounter{equation}{0}  % reset counter 

In this appendix we present the explicit expressions for the
 amplitudes depicted in Table \ref{results}.  Note, in accordance with
 the definition used for dispersive corrections as well as the power
 counting, we only keep those amplitudes that contain two--nucleon
 cuts in time--ordered perturbation theory (TOPT). Especially, we
 dropped the so--called streched boxes.

Using the same labels as in the table, one finds for the corresponding
corrections to the $\pi d$ scattering length

\be
 {a_{\pi d}^{disp}}= -\frac{g_A^2m_{\pi}^2 }{6\pi f_{\pi}^6(1{+}m_{\pi}/2M_N)(1{+}q_0/2M_N)}\int \frac{d^3q}{(2\pi)^3}
\frac{q^2(\alpha{+}\beta)^2}{q_0-q^2/M_N+i\epsilon}\;
\label{explexpr}
\ee
where
\be
\alpha=\left (I_1(q)-\frac{3}{2\sqrt{2}}I_2(q) \right) \ \mbox{and} \
\beta =\frac{f_{\pi}^2}{M_N}\left (u(q)+\frac{w(q)}{\sqrt{2}}\right).
\ee
Here $I_1$ and $I_2$ are the integrals that correspond to the overlap of the deuteron wave function 
($u(q)$ and $w(q)$ for the S- and D-waves,  respectively) with the one pion exchange operator
\be
\nonumber
I_1(q)&=&-\int \frac{d^3p}{(2\pi)^3} \frac{(1+({\vec p}\cdot{\vec q})/q^2)}{2\omega_{{\vec p}+{\vec q}}} 
\left (u(p)+\frac{w(p)}{\sqrt{2}}\right)\left(\frac{1}{P_1} + \frac{1}{P_2}\right),\\
I_2(q)&=&-\int \frac{d^3p}{(2\pi)^3} %\frac{(1-\frac{({\vec p}\cdot{\vec q})^2}{p^2q^2})}
\frac{(1-({\vec p}\cdot{\vec q})^2/(p^2q^2))}
{2\omega_{{\vec p}+{\vec q}}} w(p)
\left(\frac{1}{P_1} + \frac{1}{P_2}\right)
\ee
where $P_1$  and $P_2$ correspond to the TOPT components of the pion propagator
$
P_1=q_0-\omega_{{\vec p}+{\vec q}}-(p^2+q^2)/2M_N$ and 
$P_2=-\omega_{{\vec p}+{\vec q}}-(p^2+q^2)/2M_N
$
with $\omega_{\vec q}=\sqrt{{\vec q}\:^2+m_{\pi}^2}$.

The diagrams of  Table \ref{results} can be easily matched
to the individual terms of Eq.~(\ref{explexpr}): type 1 contains $\alpha^2$,
type 2 contains $\beta^2$, whereas the interference terms of type 3 contain
$2\alpha \beta$.
For the direct terms, labled as $d$ in the table, one needs to take
$q_0=m_\pi$ and for the crossed terms, labled as $c$ in the table,
 $q_0=-m_\pi$.  All necessary information on how the $NN$ interaction is
included in the intermediate state can be found in the Appendix of 
Ref. \cite{recoils2}.

All integrals are evaluated up to a sharp momentum cut--off of 1
GeV. All higher momentum components are to be absorbed in a counter
term that is to be included at order $\chi^2$ (c.f. discussion in
section 2). By enlarging the cut--off by a factor of 20 we checked that
the integrals change by less than 10 \% --- fully
in line with the power counting.

\end{document}